\documentclass[aps,prb,groupedaddress,reprint,showpacs]{revtex4-1}
\usepackage{graphicx}
\usepackage{bm}
\usepackage{color}
\begin{document}
\title{
  Synergy of Binary Substitutions for Improving the Cycle Performance in LiNiO$_2$
  Revealed by Ab Initio Materials Informatics
}
\author{Tomohiro Yoshida$^{1}$}
\author{Ryo Maezono$^{2,3}$}
\author{Kenta Hongo$^{4,3,5,6}$}

\affiliation{$^{1}$
  Department of Computer-Aided Engineering and Development, 
  Sumitomo Metal Mining Co., Ltd., 3-5, Sobiraki-cho, Niihama, Ehime 792-0001, Japan
}

\affiliation{$^{2}$
  School of Information Science, JAIST, Asahidai 1-1, Nomi, Ishikawa, 923-1292, Japan
}

\affiliation{$^{3}$
  Computational Engineering Applications Unit, RIKEN, 2-1 Hirosawa, Wako, Saitama 351-0198, Japan
}

\affiliation{$^{4}$
  Research Center for Advanced Computing Infrastructure, 
  JAIST, Asahidai 1-1, Nomi, Ishikawa 923-1292, Japan
}

\affiliation{$^{5}$
  Center for Materials Research by Information Integration, 
  Research and Services Division of Materials Data and Integrated System, 
  National Institute for Materials Science, 1-2-1 Sengen, Tsukuba, Ibaraki 305-0047, Japan
}

\affiliation{$^{6}$
  PRESTO, Japan Science and Technology Agency, 4-1-8 Honcho, Kawaguchi-shi, Saitama 322-0012, Japan
}

\date{\today}

\begin{abstract}
  We explore LiNiO$_2$-based cathode materials with
  two-element substitutions
  by an ab initio simulation based materials informatics (AIMI) approach.
  According to our previous study,
  a higher cycle performance strongly correlates with 
  less structural change during charge--discharge cycles;
  the latter can be used for evaluating the former.
  However, if we target the full substitution space,
  full simulations are infeasible even for all binary combinations.
  To circumvent such an exhaustive search,
  we rely on Bayesian optimization.
  Actually, by searching only 4\% of all the combinations,
  our AIMI approach discovered two promising combinations, Cr-Mg and Cr-Re,
  whereas each atom itself never improved the performance.
  We conclude that the synergy never emerges 
  from a common strategy restricted to
  combinations of ``good'' elements that individually improve the performance.
  In addition, we propose a guideline for the binary substitutions
  by elucidating the mechanism of crystal structure change.
  
\end{abstract}
\maketitle
\section{Introduction}
Lithium-ion batteries (LIBs) have a high-voltage and high-capacity and
hence are widely used as secondary batteries for mobile devices
and hybrid/electric cars.~\cite{Kang, Manthiram}
The cathode material in LIBs is one of the most important factors 
that determine the battery performance.
Thus, its development has recently attracted much attention
from the viewpoint of industrial applications.
A practical solution for improving the performance is
``atomic substitution.''~\cite{Guilmarda,Guilmardb,Pouillerie,Sathiyamoorthi,Kondo,Cho,Delmas,Saadoune,Rossen,Venkatraman,Mohanb,Kwon,Kim,Yoona,Yoonb,Kimb}
For example, consider LiNiO$_2$ (LNO)~\cite{Myung,Dahn,Ohzuku},
which is known to have a higher capacity and lower cost
than LiCoO$_2$ (LCO). 
However, LNO has fewer cycle characteristics than LCO.
Indeed, the substitution of Ni sites in LNO with Co and Al,
Li(NiCoAl)$_2$,
-- a commercially used cathode material -- 
has prolonged the cycle life compared to LNO itself.
Note that Co and Al are respectively known to improve
the rate performance and thermal stability.~\cite{Ohzuku3,Ohzuku4,Lee2,Schipper01012017}
This indicates that ``individually good'' elements 
come together to tune the corresponding characteristics.
This is a commonly used strategy for improving
battery performances.
A situation may occur, however, where a combination of ``no-good'' elements
has the potential of improving the performance.
But it is really unclear if such a synergistic effect emerges.
Unfortunately, however, 
we cannot straightforwardly establish 
an experimental verification 
even for binary substitutions having
a huge number of possible combinations.
This is because such an exhaustive search based on experiments
requires enormous amounts of time and high costs for the synthesis of
candidate materials.

\par
One of the most promising solutions to the above
exhaustive search problem is materials informatics 
(MI)~\cite{2017RAM,Agrawal,Poryrailo}
-- a recently emerging paradigm in materials science
combined with information and data science.
There are a number of successful MI studies that have explored
new materials with desirable 
properties.~\cite{Seko,Hautier,2017IKE,2019WU,Aykol2,Ceder,Nishijima}
Since a sufficiently large amount of experimental data on material
properties is generally unavailable unlike in other research areas 
(e.g., Bioinformatics),
computational approaches are useful for generating the data.
In most cases regarding electronic properties,
ab initio simulations can satisfactorily generate data 
to construct machine learning-based prediction models,
thereby tackling the exhaustive search problem.
This may be called ``ab initio Materials Informatics'' (AIMI),
though computationally proposed candidates 
should be verified experimentally.
In battery materials, AIMI approaches have been used
successfully to explore cathode coating materials~\cite{Aykol2},
and new cathodes with a better 
capacity and thermal stability~\cite{Ceder}.

\par
Focusing on the cycle performance of LIBs, an AIMI approach
based on a high throughput screening combined with
density functional theory (DFT) calculations has
been applied to co-substituted LiFePO$_4$ cathode materials, and it
successfully found element combinations that could prolong
the cycle characteristics.~\cite{Nishijima}
The volume change during the charge--discharge cycle
and planer mismatch adopted in their study 
worked as the screening criterion.
This success can be attributed to the fact that
capacity fading is caused by 
microcracks during charge--discharge cycles~\cite{Yoon},
though ab initio simulations cannot directly 
evaluate the cycle performance.
Very recently, we have investigated 
which unary substitutions improve
the LNO cathode by computational screening.~\cite{Yoshida}
Our findings are as follows: (i) 
DFT simulations must incorporate
van der Waals (vdW) corrections in
exchange--correlation functionals adopted
in order to describe the structural changes
during electrochemical cycling;
(ii) $c$-axis contraction
is more essential for the cycle characteristics;
(iii) our descriptor analysis based on
sparse modeling elucidates important descriptors
correlating with the contraction.
Finally, our AIMI approach indicated that
Nb is the most promising candidate for
the substitute element.

\par
Our previous study provides the basis for the further
exploration of new cathode materials beyond unary substitutions.
As we mentioned before, however, binary substitutions
have a much larger search space, so we employ
a Bayesian optimization technique~\cite{Jones,Shahriari}
to efficiently discover the best binary combinations
from the huge search space.
In this study, we efficiently found synergetic binary substitutions
by computing only about 4\% of the search space,
and we elucidate the mechanism of structural changes in a MI manner,
which will be helpful for further explorations by AIMI approaches.

\section{Methodology}
\begin{figure}[tb]
  \includegraphics[width=\hsize]{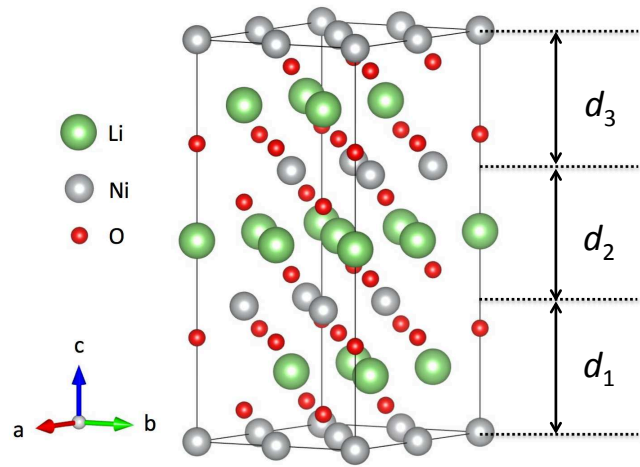}
  \caption{Crystal structure of LiNiO$_2$. 
    The super cell consists of four conventional cells.
    The structural change is measured as the 
    average over interlayer distances, $d_{i=1-3}$ (see Text).
  }
  \label{fig1}
\end{figure}

All our ab initio DFT simulations were carried out
by using the Vienna Ab initio Simulation Package
(VASP)~\cite{Kressea,Kresseb} 
with the projector augmented wave (PAW) potentials. 
The plane-wave basis set cutoff energy and k-point mesh
were set to be 650 eV and $3\times 3\times 1$, respectively,
throughout this study.
We employ the van der Waals exchange-correlation (vdW-XC)
functional~\cite{Klimesa,Klimesb} 
to accurately describe the crystal structural change 
during the charge--discharge cycle.
Fig.~\ref{fig1} shows the LNO crystal structure
with a rhombohedral R\={3}m unit cell~\cite{Ohzuku}. 
The super cell used in the study is $2\times 2\times 1$
of the conventional cell.
We assume that doping elements are dissolved in the Ni sites.
For unary substitutions, we replace Ni with $X$ (Ni:$X$=92:8);
for binary substitutions, each of the two Ni's is replaced 
by $X_1$ and $X_2$ (Ni:$X_1$:$X_2$=84:8:8),
respectively.

In order to design a high-capacity 
and high-cycle performance battery,
we must limit the candidates.
To extract approximately 80 \% of Li with only Ni redox, 
the sum of valences of $X_1$ and $X_2$ must exceed 6.
Upon the above limitation, the number of possible candidates 
amounts to 1,657 when considering
the third and lower rows in the periodic table 
up to Pa (atomic number 91) except for the
elements for which pseudo-potentials are unavailable. 
The structural changes are evaluated by the change
in the crystal structures between 0~\% and 83~\% charged states.
Although there are several possibilities for doping positions 
and a delithiated structure for each substitution,
only the most stable structure is chosen
by comparing their energies.
According to our previous study~\cite{Yoshida},
the cycle performance strongly correlates with
the $c$-axis contraction, which is measured by
the following quantity:
\begin{eqnarray}
  \Delta d_{\rm ave} = \sum^3_{i=1}\frac{|d_i^c - d_i^d|}{3}
  \label{eq1}
\end{eqnarray}
where $d^d_i$ ($d^c_i$) denotes interlayer distances
for the discharged (charged) state (see Fig.~\ref{fig1}).

We performed ab initio DFT simulations 
for all of the the unary substitutions (65 elements) 
to evaluate $\Delta d_{\rm ave}$.
In the binary substitutions, all of the possible candidates
amounted to $\sim10^5$, but in reality such a huge number of simulations
is infeasible.
To efficiently discover the best binary combination(s),
we thus adopted Bayesian optimization,
which is an experimental design algorithm associated with
Gaussian process regression.~\cite{Jones,Shahriari}
It has been reported that the design and choice of descriptors
appropriate for describing material features uniquely 
are crucial for conducting Bayesian optimization.~\cite{Seko}
The interlayer distances, $d_i ~(i = 1,2,3)$, are
determined as a balance between the 'Coulomb repulsive interaction 
between oxygen in the NiO$_6$ layer' and 'vdW attractive interactions
between the NiO$_6$' layers~\cite{Laubach,Li2,Pouillerie2,Yabuuchi,Chen-Wiegart,Basch}.
Although those interactions would be good descriptors,
it is quite hard to directly evaluate them.
Instead, as listed in Table~\ref{table1},
we collected ``elemental information of dopants'' and
``structural information of 0~\% charged state'' as 
descriptor candidates; 
all of the quantities were symmetrized 
with respect to the exchange between two dopants,
{\it e.g.}, $m_{X1}+m_{X2}$ and $m_{X1}\times m_{X2}$ 
for atomic mass $m$; we eventually selected
those quantities as descriptors by random sampling.
Based on Gaussian regression with the descriptors selected above,
we constructed a prediction model of $\Delta d_\mathrm{ave}$
and then performed Bayesian optimization 
under the maximum probability of improvement (as an acquisition function).
We employed the Common Bayesian Optimization Library 
(COMBO)~\cite{Ueno} to implement our Bayesian optimization.
\begin{table*}[tb]
  \begin{tabular}{cc} 
    \hline
    elemental information & crystal structure information \\
    \hline
    atomic number $Z$ & lattice volume $V$ \\
    atomic mass $m$ & lattice constant $a$, $b$, $c$ \\
    electronegativity $EN$ & angle formed by basic vector $\alpha$, $\beta$, $\gamma$ \\
    covalent radius $CR$ & substituted positions of $X_1$ and $X_2$ $x$, $y$, $z$ \\
    first-ionization energy $IE$ & \\
    maximum oxidation state $O$ & \\
    coefficient of van der Waals $C_6$ & \\
    \hline
  \end{tabular}
  \caption{
    List of descriptors considered in the present study.
  }  
  \label{table1}
\end{table*}

\section{Results and Discussion}

$\Delta d_\mathrm{ave}$ values for all of the unary substitutions
(65 elements) are summarized in the Appendix.
Their comparisons with LNO value 
($\Delta d_\mathrm{ave} = 0.156~\AA$)
classify the binary substitutions into 
`positive element' and `negative element' substitutions depending on
whether or not it suppresses a structural change,
{\it i.e.}, the positive/negative has
a $\Delta d_\mathrm{ave}$ value less/greater than $0.156~\AA$.
Furthermore, the negative elements can be divided
into the following two groups: the ones that dissolved into the Ni site
and the others that moved from the Ni site to the Li site 
by structural relaxation.

\begin{table}[tb]
  \begin{tabular}{c|c}
    \hline
    positive element & negative element \\
    \hline
    V,Nb,Ge,Ir,Au & Pb,Pa,Hf,Rh,Fe,Cu \\
    Al,Ti,Ta,Mg,Ga & Po,W,Zn,Cr,Zr,Pd \\
    Ru,Bi,Dy,Os,Sb & Mo,Ca,Yb,Cd,Ag \\
    Mn,Tc,Y,Tb,Re & Gd$^\ast$,Ce$^\ast$,Eu$^\ast$,Tl$^\ast$,Pm$^\ast$,Sm$^\ast$ \\
    Tm,Lu,Pt,Ho,In & Hg$^\ast$,Na$^\ast$,Th$^\ast$,Nd$^\ast$,Pr$^\ast$,Sr$^\ast$ \\
    Er,Sc,Sn & La$^\ast$,Ac$^\ast$,Ba$^\ast$,Ra$^\ast$,K$^\ast$,Rb$^\ast$ \\
     & Fr$^\ast$,Cs$^\ast$ \\
    \hline
  \end{tabular}
  \caption{
    Classification of presence/absence of the
    $\Delta d_\mathrm{ave}$ suppression effect 
    in unary substitutions. 
    In the negative element, 
    the element marked with $\ast$ moves 
    from the Ni site to the Li site.
  }  
  \label{table2}
\end{table}

V is the best dopant among the unary substitutions,
having $\Delta d_{\rm ave}=0.124 \AA$.
We note that Zr and Na are classified into negative elements
by the ab initio simulations,
but these elements are known to improve cycle performance as follows:
Zr stabilizes the cation ordering of Li and Ni~\cite{Yoona};
Na stabilizes solid solutions at the Li site 
(pillar effect)~\cite{Kimb}.
These effects are beyond the scope of the present simulations,
so we no longer consider these elements, which prevents
us from obtaining inconsistent results.

As mentioned above, 
V is the best element for the unary substitutions.
Here, we investigate if binary substitutions 
further suppress $\Delta d_\mathrm{ave}$ 
by using Bayesian optimization.~\cite{Jones,Shahriari}
First, we randomly select 31 candidates 
and evaluate their $\Delta d_\mathrm{ave}$ values
by ab initio DFT simulations.
Next, by using the descriptors shown in Table~\ref{table1},
we construct a prediction model of $\Delta d_\mathrm{ave}$
based on the Gaussian process regression learned from the data set.
We finally conduct Bayesian optimization for exploring
the best binary combinations.

\par
\begin{figure}[tb]
  \includegraphics[width=\hsize]{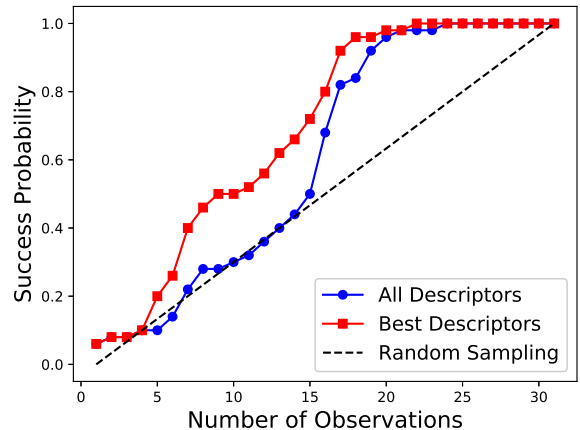}
  \caption{
    Success probability for the Bayesian optimization 
    with all of the descriptors (blue) and
    the best descriptors (red). 
    Random sampling (black) is also shown for comparison.
  }
  \label{fig2}
\end{figure}

Figure~\ref{fig2} shows how quickly the Bayesian optimization
and random search found the smallest $\Delta d_\mathrm{ave}$.
For Bayesian optimization, we considered two sets of descriptors
including all and the ``best'' selected ones (explained later).
The average number of observations ($N_\mathrm{ave}$) 
required for finding the optimized solution 
using Bayesian optimization with all of the descriptors 
and the random search were 13 and 15, respectively. 
As can be seen in Fig.~\ref{fig2},
the Bayesian optimization rapidly 
reaches the smallest $\Delta d_{\rm ave}$
as compared with the random search,
but its efficiency is not so high.
In particular, when the number of observations is less than 15,
the success probability of Bayesian optimization
with a set of all of the descriptors
is comparable to that of random sampling.
This is probably because the set of all of the descriptors 
shown in Table~\ref{table1}
is inefficient for predicting $\Delta d_{\rm ave}$ accurately.
Therefore, as explained below, we conducted the descriptor analysis by
LASSO regression~\cite{Tibshirani}
and then extracted the most important ones
in terms of the $\Delta d_{\rm ve}$ prediction;
we anticipate this reduced set of descriptors
can improve the efficiency of Bayesian optimization.

In our descriptor analysis, we first divide the data 
into training/test sets with the ratio of 8/2.
The prediction model of $\Delta d_{\rm ave}$ is 
based on the LASSO regression learned from the training set.
With the use of this trained model, 
we predict $\Delta d_{\rm ave}$ values for the test set and 
compare them to the corresponding ab initio values.
Its prediction accuracy is evaluated
in terms of the mean square error.
The LASSO is known to automatically select 
important descriptors for prediction.
As a result, we obtained six descriptors 
($V$, $\alpha$, $m_{X1}\times m_{X2}$, 
$EN_{X1}\times EN_{X2}$, $a$, $z_{X1}\times z_{X2}$),
and by using them, Bayesian optimization was implemented again,
which is denoted by the ``best descriptors''.
As expected, $N_{\rm ave}$ was 10 for the best descriptors,
and the success probability improved (see Fig.~\ref{fig2}).
We thus expect that these descriptors are efficient 
even with further searches, 
and then, we proceed with the Bayesian optimization 
for the remaining 1,626 candidates.

\par
\begin{figure}[tb]
  \includegraphics[width=\hsize]{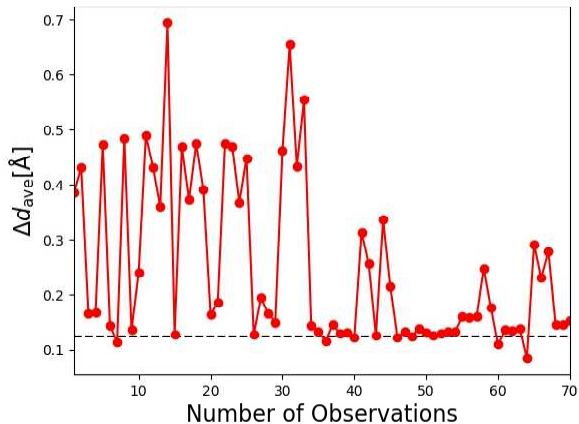}
  \caption{
    Search history for the binary substitutions.
    Black-dashed line shows 
    the smallest $\Delta d_{\rm ave}$ 
    for the unary substitutions.
  }
  \label{fig3}
\end{figure}

Thirty nine times observations for the Bayesian optimization
with the best descriptors were carried out after 
31 times observations for random sampling,
where their search history is shown in Figure~\ref{fig3}.
Meanwhile, the random sampling (observation \#: 1$\sim$31) 
chooses $\Delta d_{\rm ave}$ values distributed
from large to small values, and
the Bayesian optimization (32$\sim$70) 
selects smaller $\Delta d_{\rm ave}$ values
(N.B., our acquisition function is the maximum probability
of improvement as described 
in the Methodology section.

From the above observations by the Bayesian optimization, 
we discovered V-Ga, Ti-Fe, Al-Mn, Al-Cr, Mg-Cr, and Cr-Re 
as the combinations that suppress $\Delta d_{\rm ave}$ 
better than the best unary substitution.
Although V-Ga and Al-Mn are composed of positive elements,
the other combinations include negative elements, Fe or Cr.
Nevertheless, for the latter, 
their $\Delta d_\mathrm{ave}$ values showed more suppression
as compared to the best unary substitution,
which can be interpreted as being synergistic effects.

\begin{figure*}[tb]
  \begin{center}
    \begin{tabular}{c}
      \begin{minipage}{0.5\hsize}
        \includegraphics[width=\hsize]{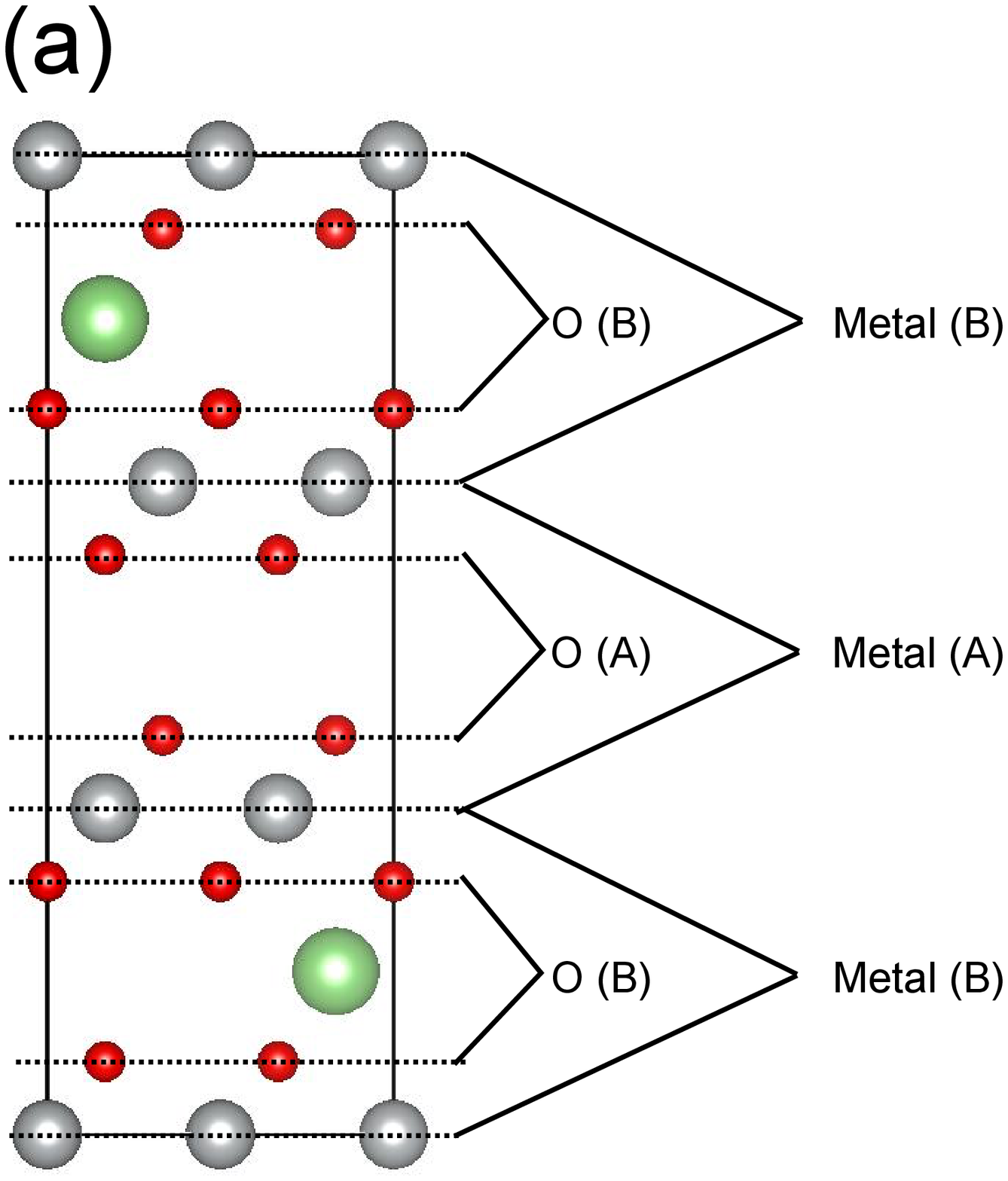}
      \end{minipage}
      \begin{minipage}{0.5\hsize}
        \includegraphics[width=\hsize]{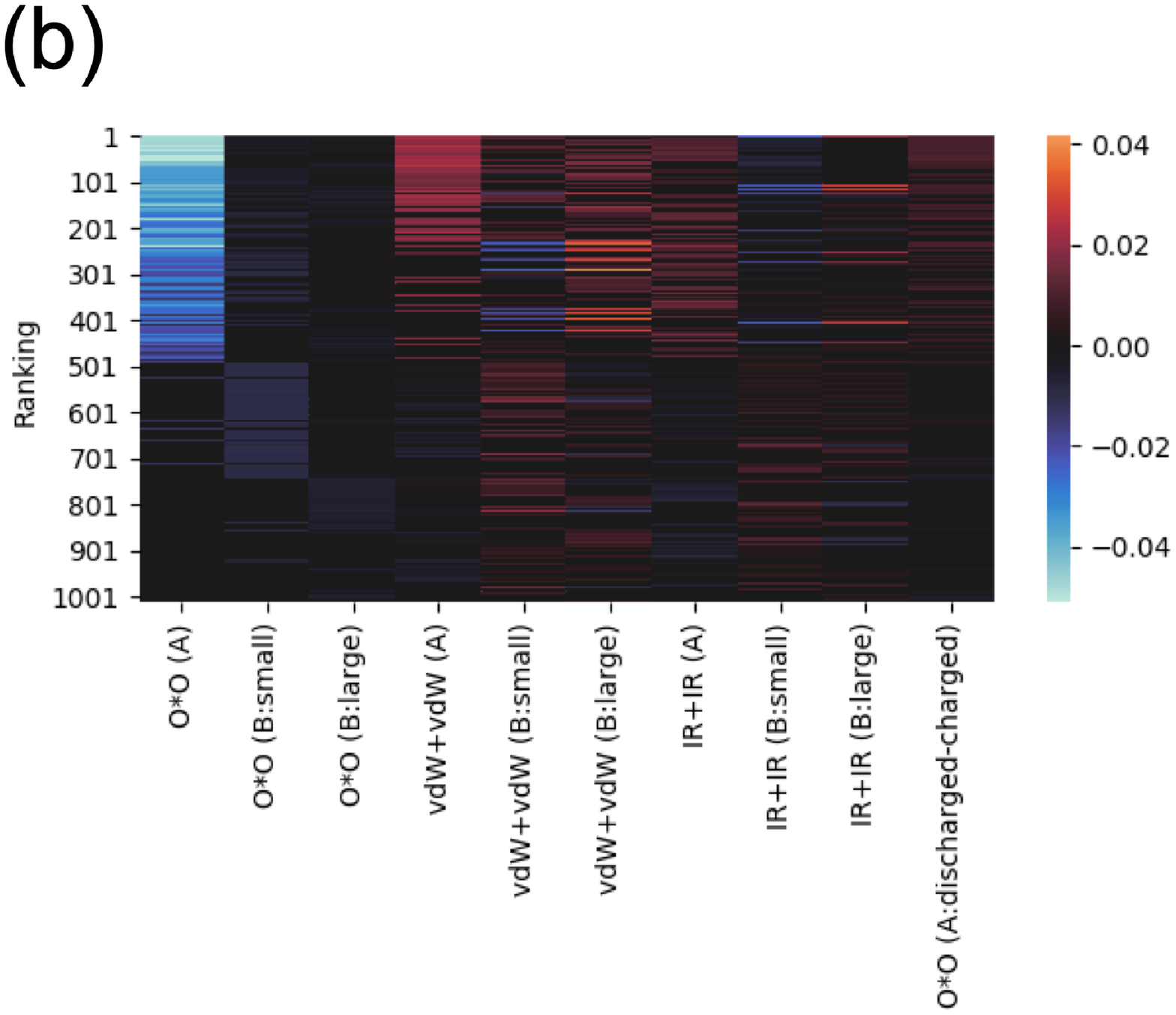}
      \end{minipage}
    \end{tabular}
    \caption{
      (a) Descriptor classification. 
      (A) Ni-O slabs sandwiching a Li-discharged layer 
      and (B) Ni-O slabs sandwiching a Li-charged layer.
      (b) Heat map of regression coefficients 
      arranged according to the models'
      coefficient of determination R$^2$.
      The largest R$^2$ was 0.77.
      From left, ``a product of Badar charge on oxygen 
      (A, B small, B large),''
      ``a sum of vdW coefficients in the Ni-O slab 
      (A, B small, B large),'' 
      ``a sum of ionic radii of cations (A, B small, B large),''
      and ``a difference in Bader charge on oxygen 
      between charged and discharge states.''
      See Text for more details.
    }
    \label{fig4}
  \end{center}
\end{figure*}

In order to elucidate why the synergy emerges, 
the suppression mechanism in the binary substitutions
is investigated here.
We construct a regression model of $\Delta d_{\rm ave}$ predictions
learned from 27 combinations 
where dopants do not move to the Li sites.
Changes in interlayer distances $d$ during cycling
are anticipated to be determined 
by the ``Coulomb repulsive interaction 
between oxygen in the NiO$_6$ layer,'' ``vdW attractive interactions
between the NiO$_6$ layers,'' ``ionic radii of doping cations,'' etc.
Accordingly, the following quantities are considered as
our descriptors entering the regression model:
``a product of the averaged Bader charges~\cite{Bader,Tang}
over the facing oxygen layers,''
``a sum of the averaged vdW coefficients over the facing Ni-O slabs,''
``a sum of the averaged ionic radii over the facing Ni layers,''
and ``the difference in the averaged Bader charges 
within the oxygen layer between charged and discharged states.''
In total, we obtained 10 descriptors for 
the $\Delta d_\mathrm{ave}$ prediction.
At the charge rate used in this study, 
Ni-O slabs can be classified into the following two cases 
[see Fig.~\ref{fig4} (a)]:
(A) Ni-O slabs sandwiching a Li-discharged layer 
and (B) Ni-O slabs sandwiching a Li-charged layer. 
The above descriptors are calculated for each case.
Note that two independent O layers exist in the (B) Ni-O slab
and the corresponding descriptors 
distinguished by their magnitudes 
are independently treated.

We consider all possible combinations of 10 descriptors
to construct the regression models.
Figure~\ref{fig4} (b) shows their regression coefficients 
in a heat map form.
Looking at models with higher R$^2$ 
(coefficient of determination) values, 
it can be seen that 
the charged product of the oxygen layer 
(A)/the sum of the vdW coefficient has 
a negative/positive contribution.
This is because the larger Coulomb repulsion 
and smaller vdW attraction suppress 
a shrinkage in the $c$-axis direction.
These two descriptors are the most important factors 
describing the change along the $c$-axis.
In other words, in order to improve the cycle characteristics, 
it is preferable to substitute Ni atoms with those
which have a low vdW coefficient, 
holding large Bader charges on oxygen layers.
We also found that a large change in the Bader charge on oxygen 
accompanying the charge--discharge cycle significantly
contributes to the suppression of structural change.
This is because
a smaller change in the electronic structure 
during the charge/discharge cycle suppresses the structural change.
Although the ionic radius of the cation is less important than
the product of the oxygen charge and sum of the vdW coefficient,
it is required for constructing an accurate model.
\par
It should be noted that cations substituted with Ni require
a small ionic radius in (A) and inversely a large one in (B).
Consequently, any unary substitution cannot satisfy
both the requirements simultaneously, and hence,
its $\Delta d_{\rm ave}$ suppression is smaller than
that of the binary substitution.
In other words, these inverse requirements are 
the origin of the synergistic effects.
Indeed, Cr-Re ($\Delta d_\mathrm{ave}=0.08 \AA$) shows 
the most suppression of $\Delta d_\mathrm{ave}$ change,
where a large Cr cation and a small Re cation are
doped into (B) and (A), respectively.
It is concluded that the understanding of
why synergistic effects occur is important
in order to design the best performance material.

\section{Conclusions}
We investigated unary and binary substitutions in LiNiO$_2$ 
to improve its cycle performance.
Owing to their strong correlation,
the cycle characteristics were evaluated from the perspective of
structural change along the $c$-axis during cycling,
and this was investigated for various doping elements
by using ab initio simulations.
For the unary substitutions, 
an exhaustive search based on the ab initio simulations 
revealed that V is the best for a unary substitution.
As for the binary substitutions, our Bayesian optimization
explored the candidate combinations more efficiently
than random sampling, and results revealed synergistic effects
for combinations of doping elements that never improved
the cycle performance individually (Cr-Mg and Cr-Re).
We also proposed guidelines for battery material design.
Our ab initio materials informatics approach
presented here can be regarded as 
a promising fundamental technique 
for further exploration of high-performance battery materials.


\section{Acknowledgments}
The computations in this work were performed by 
using the facilities of RCACI 
(Research Center for Advanced Computing Infrastructure) at JAIST. 
T.~Y. would like to thank T.~Kosasa, 
K.~Ryoshi, T.~Toma, and S.~Yoshio 
for their fruitful discussions and technical support. 
K.~H. is grateful for financial support 
from KAKENHI grants (17K17762 and 19K05029), 
a Grant-in-Aid for Scientific Research 
on Innovative Areas (16H06439 and 19H05169),
the FLAG-SHIP2020 project 
(MEXT for the computational resources, projects hp180206 and 
hp180175 at K-computer), and
PRESTO (JPMJPR16NA) and the Materials research
by Information Integration Initiative (MI$^2$I) project of 
the Support Program for Starting Up Innovation Hub
from the Japan Science and Technology Agency (JST).
R.~M. is grateful for financial support 
from MEXT-KAKENHI (project JP16KK0097), 
the FLAG-SHIP2020 project 
(MEXT for the computational resources, projects hp180206 
and hp180175 at K-computer),
and the Air Force Office of Scientific Research 
(AFOSR-AOARD/FA2386-17-1-4049).
\section{appendix}
We sumarize the calculated $\Delta d_{\rm ave}$ for unary substitution in Table \ref{tableS1}.
\begin{table*}[htb]
  \begin{tabular}{lcccccc} 
    \hline
    Element & $\Delta d_{\rm ave}$(\AA) & Element & $\Delta d_{\rm ave}$(\AA) & Element & $\Delta d_{\rm ave}$(\AA) \\
    \hline
    V  & 0.124 & Pt & 0.147 & Ag & 0.199 \\
    Nb & 0.125 & Ho & 0.148 & Gd & 0.216 \\
    Ge & 0.132 & In & 0.149 & Ce & 0.257 \\
    Ir & 0.134 & Er & 0.149 & Eu & 0.283 \\
    Au & 0.134 & Sc & 0.150 & Tl & 0.290 \\
    Al & 0.135 & Sn & 0.155 & Pm & 0.306 \\
    Ti & 0.135 & Pb & 0.159 & Sm & 0.314 \\
    Ta & 0.135 & Pa & 0.160 & Hg & 0.337 \\
    Mg & 0.135 & Hf & 0.160 & Na & 0.347 \\
    Ga & 0.137 & Rh & 0.161 & Th & 0.385 \\
    Ru & 0.137 & Fe & 0.161 & Nd & 0.395 \\
    Bi & 0.139 & Cu & 0.162 & Pr & 0.420 \\
    Dy & 0.139 & Po & 0.163 & Sr & 0.426 \\
    Os & 0.140 & W  & 0.165 & La & 0.440 \\
    Sb & 0.140 & Zn & 0.167 & Ac & 0.493 \\
    Mn & 0.141 & Cr & 0.167 & Ba & 0.574 \\
    Tc & 0.141 & Zr & 0.169 & Ra & 0.622 \\
    Y  & 0.142 & Pd & 0.172 & K  & 0.685 \\
    Tb & 0.143 & Mo & 0.177 & Rb & 0.777 \\
    Re & 0.144 & Ca & 0.189 & Fr & 0.856 \\
    Tm & 0.146 & Yb & 0.191 & Cs & 0.865 \\
    Lu & 0.147 & Cd & 0.193 & \\
    \hline
  \end{tabular}
  \caption{$\Delta d_{\rm ave}$ for unary substitution.
  }
  \label{tableS1}
\end{table*}
\bibliography{references}
\end{document}